# Towards the mechanism and high performance of solid-state Li batteries

*L.J. Zhang*

**Lithium ion batteries have played an important role in supporting diverse energy storage scenarios in practical commercial applications**[1-6]**, such as portable electronic devices and electric vehicles (EV)**[7-10]**, et al. With the increasing demand for higher safety and energy density of energy storage batteries, solid-state lithium batteries (SSLBs) using solid-state electrolytes have become a focus of research for meeting high safety and performance needs due to their potential higher energy density and avoidance of flammable liquid electrolytes**[1, 11-17]**. However, some critical problems and challenges have been exposed, hindering the development and practical application of SSLBs, such as the low room temperature ionic conductivity of solid electrolyte, the risk of short circuit caused by lithium dendrite piercing the electrolyte, the damaging thermal runaway at 300 ℃ and above, the volume expansion and structural degradation of electrodes, the poor contact or contact degradation of electrode/electrolyte interface, the increase in internal resistance that hinders ion migration resulting from the formation of interphase, etc. In order to address these challenges, it's essential to obtain in-depth insights into mechanisms and systematic optimization of SSLBs, including interfaces, electrolytes, and battery structures. Here, this minireview provides a brief summary, including strategies for electrode and electrolyte preparation, advanced battery characterization techniques, and the latest computational and simulation methods to advance understanding of kinetic or atomic scale mechanisms. The above contents will play an active role in promoting the development and practical application of safer and higher performance SSLBs.**

**1 The preparation strategy**

The composition and structural design of electrode and electrolyte materials are important for battery performance, and it should be noted that preparation strategies also play a crucial role. During the preparation of electrodes and electrolytes, conventional methods usually employ time- and energy-intensive process such as milling, pressing and heat treatment process varying from hundreds to over one thousand degrees C for several hours or days, and vacuum or Ar\N$_2$ conditions are also frequently employed to avoid the interaction between electrode\electrolyte and O$_2$\water[18-20]. Those fast and facial, free of toxic reagents methods have attracted increasing interests because of the merits of higher potential for cost-effective production. Spark plasma sintering (SPS) can serve as a fast heat treatment tool for solid electrolyte, which can complete the process within 10 min[21]. However, high temperature of about 600 °C and high pressure are still required for the SPS method in some research. The diffusion-dependent electrode fabricated by Lee JJ et al only need heat treatment at 80-100 °C[22], which provides inspiration for the significant energy-saving effect of the preparation process. It's worth mentioning that recent research has shown a microwave-assisted quick-making method for disordered rocksalt oxides (DRX) cathode, which can be completed within 5min at ambient air[23].

The use of excessive lithium is a commonly used preparation strategy, for instance, high temperature heat treatment can lead to lithium loss, so an excess of lithium (e.g. an excess of 10%) is usually required before heat treatment to compensate for the loss[24, 25]. In some research, in order to achieve sufficient lithiation, five times excessive lithium bis(trifluoromethylsulfonyl)imide (LiTFSI) was used in the process of preparing electrolytes[26]. It's also reported that adding excess Li can improve the grain-boundary ionic conductivity of some electrolyte such as lithium aluminum germanium phosphate (LAGP)[21].

It is worth noting that some samples should be carefully stored in an oxygen free and moisture free environment before assembly, which can be achieved by preheating vacuum quartz ampoule bottles or glovebox with an argon atmosphere ($O_2$< 0.1, 0.5 or 1 ppm, water < 0.1, 0.5 or 5 ppm)[26]. The specific storage conditions usually depend on the sensitivity of the material to moisture and oxygen.

Additives should be carefully selected to improve the performance of solid-state cell. For example, carbon black can be used as a conductive additive for cathodes[27], electrolyte can also act as one composition of cathode complex[28].

Structure optimization, particle size distribution and even pressure control are all critical for solid electrolytes, such as $Li_7SiPS_8$[29]. Research results have demonstrated that stack pressure could control the impedance of the particle-particle contacts in solid electrolytes.

In addition, thin film or coatings have been reported to improve the performance of electrode stability. For example, nanofilms on the surface of solid-state electrolyte have been proved to be effective to enhance the contact between electrodes and electrolyte and inhibit lithium dendrite propagation[28]. It's also reported that only 2-4 nm thin cationic polymer coating for NCM electrode via a spray dryer enhanced the interface stability between NCM and the $Li_6PS_5Cl$ electrolyte[30].

## 2 Advanced battery characterization techniques

Characterization techniques are essential for advanced insights of solid-state batteries, such as AFM, scanning electron microscopy (SEM), focused-ion beam scanning electron microscopy(FIB-SEM), TEM, high-angle annular dark-field scanning transmission electron microscopy (HAADF-STEM), X-ray photoelectron spectroscopy (XPS), X-ray CT, X-ray diffraction (XRD), synchrotron X-ray analyses, hard and soft X-ray absorption spectroscopy (XAS), Fourier transform infrared (FTIR), Raman spectroscopy, electrochemical impedance spectroscopy (EIS), electron energy loss spectroscopy measurement (EELS), secondary ion mass spectroscopy (SIMS), inductively coupled plasma spectroscopy(ICP), solid-state nuclear magnetic resonance spectroscopy(ssNMR), neutron imaging, et al [27, 31-41].

The HAADF-STEM is not suitable for imaging light elements due to the rapidly decaying intensity with the number of atoms, so annular bright-field (ABF) STEM is usually employed for those light elements (e.g. Li)[32].

FIB-SEM is an efficient tool for cross-section observation for coated and pristine reactive materials and electrode composite, where ion beam etching can be used to cut samples at specific analytical positions to expose cross-sectional areas, thereby achieving high-resolution imaging based on SEM.

Chemical compositions of materials can be determined via ICP, but for the molar amount determination of some elements (e.g. Li, F) in potential impurity phases, some complimentary technique should be considered, such as ssNMR[23].

Neutron imaging via NeXT (Neutron and X-ray Tomography) instrument has been reported as a tool to visualize spatial distribution changes of the lithium during battery operation, which is achieved by detection of neutron attenuation along the axis of battery upon cycling[27]. For the accuracy of quantitative testing results of neutron imaging, low neutron absorption materials such as aluminum, should be considered as the casing, if the battery should have one. Isolation of air and moisture for battery stacks and the separation between battery stacks and metal casting are all necessary procedures, which usually use O-ring seals and polyimide sheath electrical insulation materials, respectively. Moreover, some pure $^6Li$ should be added to anodes before the cell assembly.

The characterization of the thin solid electrolyte interphase (SEI) with indistinguishable boundaries is difficult, but SEI can affect the cell capacity, cycling stability and so on[31]. So usually a combination of multiple characterize techniques, computation and simulation are required for SEI evolution.

## 3 Computational and simulation methods

Various computational and simulation methods have been employed in the research of solid batteries, mainly including ab initio molecular dynamics simulation (AIMD), molecular dynamics (MD), reactive forcefield (ReaxFF), eReaxFF, finite element analysis, first principle calculation and quantum chemistry (QC) et al[42-44].

The ab initio molecular dynamics simulation (AIMD) has been employed to obtain atomistic insights for the reaction and structure evolution of SSLBs, and many programs have been demonstrated capable for this work, such as Vienna ab initio simulation package (VASP), CP2K program. In fact, a combination of many programs and software were employed in some works, such as Gaussian 16 or CP2K for geometry optimizations, GROMACS program for the classical molecular dynamic pre-equilibrium before AIMD and RDF curves analyzation, OBGMX package for generating force field[26]. For a typical simulation, some critical aspects should be considered including PBE-GGA, exchange correlation function, cutoff energy, timestep, ensemble, simulation temperature, et al. Figure 1 shows a typical AIMD workflow and corresponding structures of electrolyte composition in literature[26].

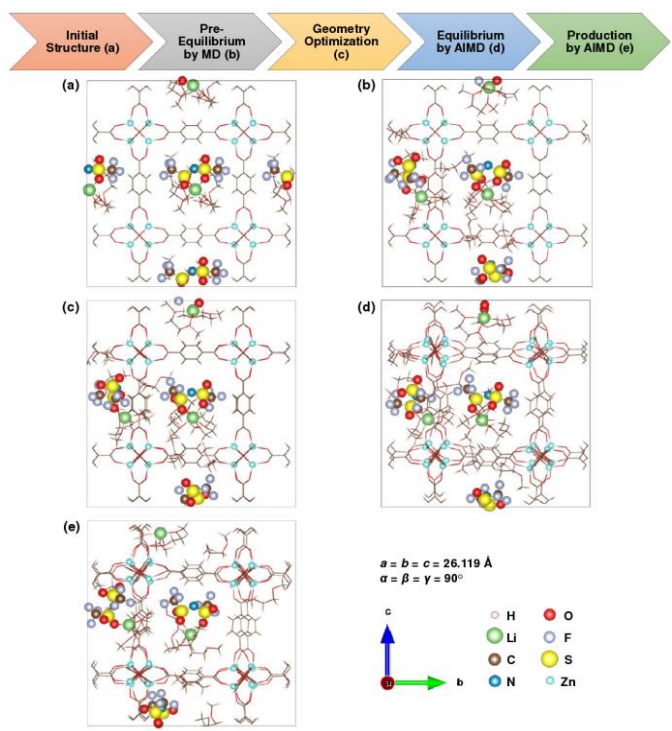

**Figure 1** AIMD workflow authors would like and corresponding structures of Li@MOF-5 (3LiTFSI + 12DME + MOF-5)[26]. (a–e) Corresponding structures of Li@MOF-5 in AIMD workflow: (a) initial structure (total 664 atoms), (b) pre-equilibrated structure by classical MD at 300 K for 2 ns, (c) optimized structure from b, (d) equilibrated structure by AIMD at 300 K for 5 ps, (e) production run by AIMD at 300 K for 50 ps. Copyright 2023 Wiley

MD simulation can be performed on a free package, large-scale atomic/molecular massively parallel simulator (LAMMPS)[45]. As an open-source classical molecular dynamics code, LAMMPS has potentials for various systems, embodying metals, semiconductors, biomolecules, polymers et al. The atomistic understanding about the lithium crystallization at the solid interfaces has been obtained recently via LAMMPS MD simulation[45]. A multi-step crystallization atomistic pathway (Figure 2) was proposed based on the MD results, which is different from the conventional understanding.

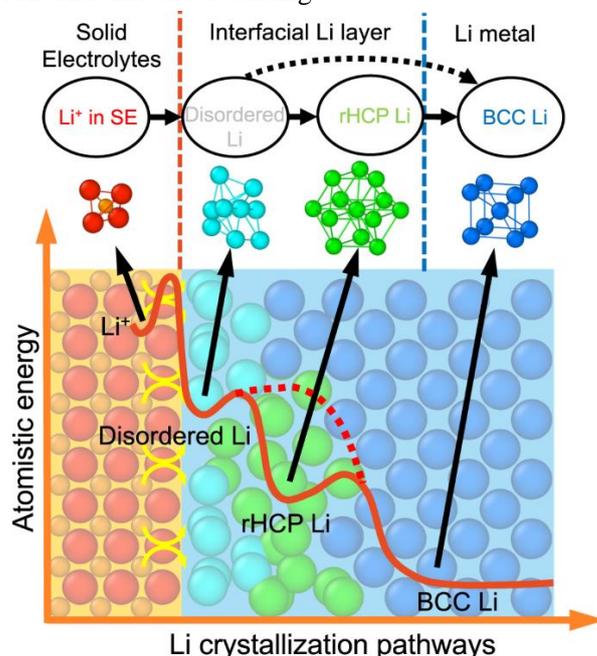

**Figure 2** A schematic of multiple-step pathways of Li crystallization. [45]. The Li+ (orange, anion shown in red) in solid electrolytes (SE) goes through disordered-Li (cyan) and/or rHCP (random hexagonal close-packed)-Li (green) in the interfacial Li layer at the SE interface, and transforms into the crystalline BCC (body-centered cubic)-Li metal (blue). Copyright 2023 Nature

The ReaxFF can describe the fracture and formation of bonds via bond-order calculation and the electronegativity-equalization method (EEM)[46-50]. For some redox processes, electron should be treated explicitly, thus eReaxFF is employed to meet this requirement.

Finite element method can also act as a valuable tool for analyzing solid-state lithium batteries, such as press distribution simulation and the diffusion of Co into the alumina[25]. For example, multiphysics analysis based on the battery design module of COMSOL software demonstrates the application of finite element method in cell[24]. Predefined related electrochemical interfaces, as well as customized chemical reactions through Nernst Planck and Butler Volmer kinetic equations in COMSOL, contribute to the simulation of lithium batteries. The isotropic volume change model can be used for simplification of simulation, even for some actual circumstance with an-isotropic volume change.

The discrete element method (DEM) has recently been employed for the analysis of powder compaction in solid electrolyte under pressure condition[29]. Finite volume (FV) simulations and the battery and electrochemistry simulation tool (BEST) can be further used to determine the effective ionic conductivities of the electrolyte in forms of pellets with the DEM results. With the help of DEM-FV simulation, the hypothesis that the change of ionic conductivity described by the activation volume serves as the main effect at variable pelletizing pressure is confirmed.